\documentclass[11pt,twoside]{article}

\usepackage{asp2006}
\usepackage{epsf}
\usepackage{psfig}
\usepackage{lscape}

\newcommand{\msun}{M$_{\odot}$}
\newcommand{\ergl}{ergs~s$^{-1}$}
\newcommand{\ergcms}{ergs~cm$^{-2}$~s$^{-1}$}
\newcommand{\cxo}{{\sl Chandra}}
\newcommand{\etal}{et al.}
\newcommand{\hii}{H\,{\sc ii}}

\newcommand{\sst}{{\sl Spitzer}}
\newcommand{\arcdeg}{\hbox{$^\circ$}}

\begin{document}
\title{Ultraluminous X-ray Sources in Interacting Galaxies}   
\author{Douglas A. Swartz}   
\affil{NASA/MSFC}    

\begin{abstract}
I give a brief review of how X-rays from nearby galaxies are
 used as direct tracers of recent star formation. 
This leads to the conclusion that it is the most luminous 
 point-like sources that are associated with star formation and
that the majority of these are high-mass X-ray binaries.
I then discuss a recent study that shows that ULXs are preferentially
 found in regions as young as or younger than typical \hii\ regions in their
 host galaxies. Finally, I describe a new study that attempts to 
determine the maximum luminosity of ULXs in the local universe by 
searching for them in interacting galaxies where the star formation rate 
is high.

\end{abstract}



\section{X-ray Sources as Tracers of Recent Star Formation} \label{s:one}

\citet{djf92} performed the first large survey of galaxies that showed
 a correlation between the total X-ray luminosity and galaxy-wide
 star-formation rate as measured by FIR ({\sl IRAS}\,) luminosities.
Although the X-ray data were not of a quality sufficient to distinguish among 
different types of X-ray sources, they argued on energetic grounds that
most of the X-ray light came from
high-mass X-ray binaries, with typical X-ray luminosities on the order of 
10$^{38}$~\ergl\ and lifetimes of 10$^5$ to 10$^6$~yr rather than
 OB-star winds and SNe (both of which are much fainter though late O stars
live longer than XRBs; see also \citet{helfand01, dalton95}).

The next major step forward awaited \cxo's superb angular resolution and broadband
(CCD) spectral senstivity to isolate individual X-ray sources in nearby galaxies.
The definitive work is \citet{ggs03}. 
They showed that the point-source X-ray luminosity functions (XLFs)
of individual normal and starburst galaxies all had roughly the same 
power law index over a wide range of source luminosities from $<$10$^{36}$~\ergl\
up to about the highest luminosities observed of $>$10$^{40}$~\ergl\
(Figure~1, left panel).
Furthermore, they showed that the overall normalization of these XLFs scaled
with the total star-formation rates (SFRs) of the individual host galaxies.
Thus, by rescaling observed XLFs to their universal XLF 
(Figure~1, right panel) one could deduce the
star formation rate of the observed galaxy.
Since the total X-ray luminosity is just the integral of the XLF, it 
follows (rather circularly back to the work of \citet{djf92}) that
the SFR can also be deduced from a measure of the total X-ray luminosity
even when individual sources cannot be resolved such as at high redshifts.

Another key result of the \citet{ggs03} study was that the 
(cumulative) XLF slope was rather flat, 
$
N(>L) \propto L^{-0.61} ,
$
where $N(>L)$ is the number of sources with luminosity greater than
$L$. This means that the most luminous sources dominate the total 
X-ray luminosity and hence the most luminous sources trace recent star-formation
activity.

\begin{figure*}[ht]
\begin{center}
\scalebox{0.3}{\includegraphics{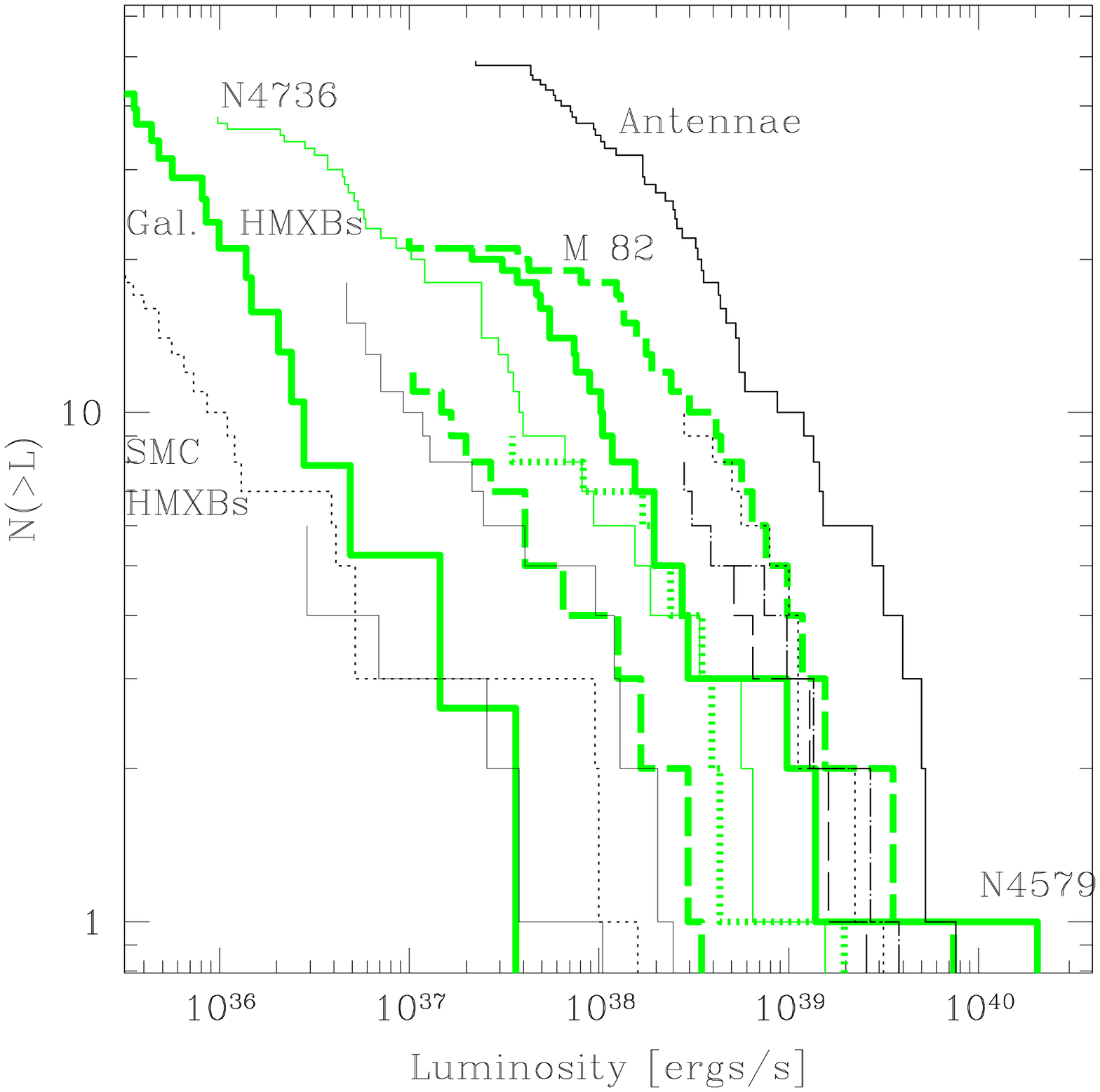}} 
\scalebox{0.3}{\includegraphics{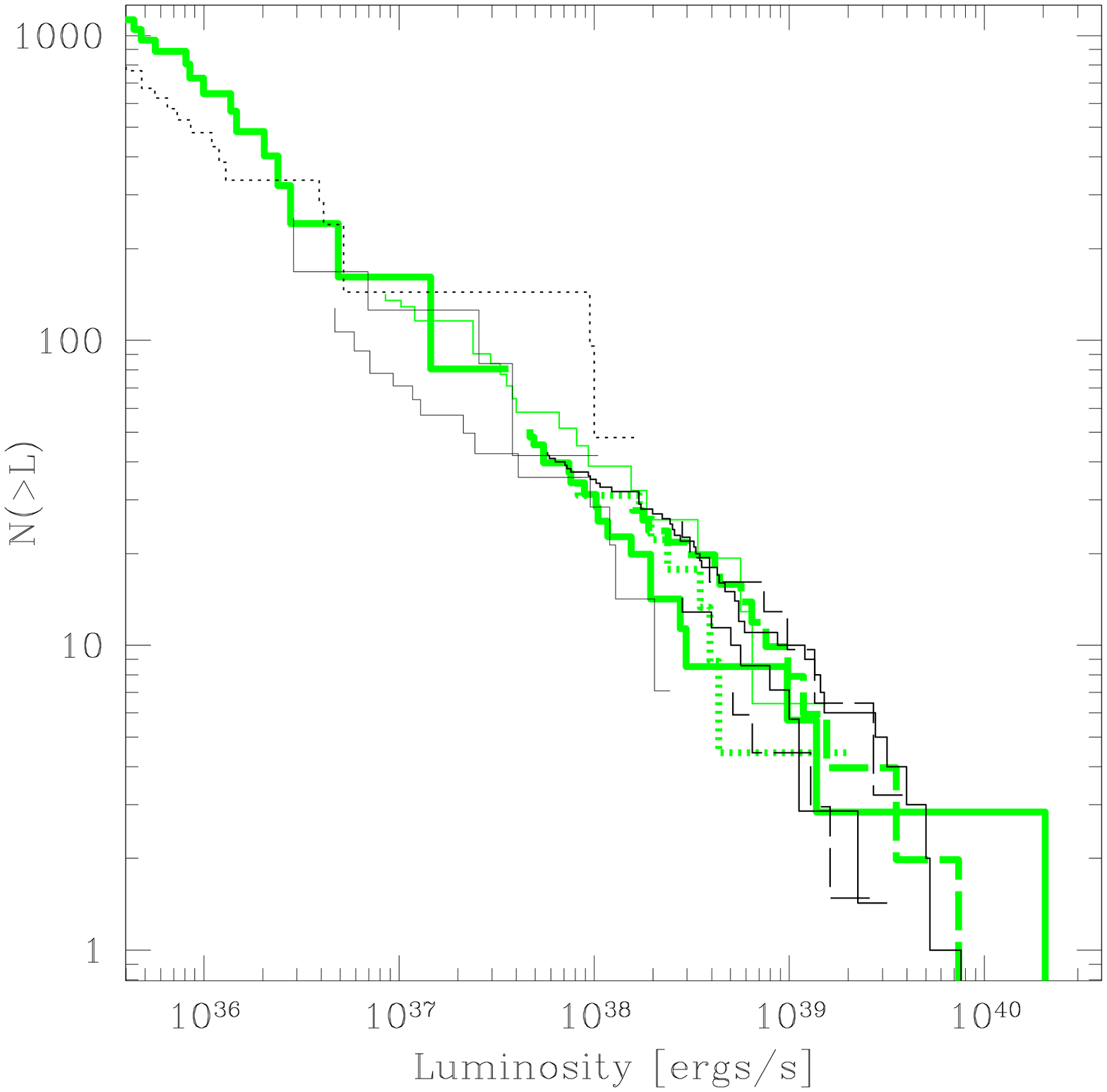}}
\end{center}
\caption[]{{\sl Left:} Cumulative XLFs for normal spiral and starburst galaxies.
{\sl Right} Same XLFs but normalized to the star-formation rate of the 
Antennae galaxies. Note that the scaled XLFs occupy only a narrow range
in $N(>L)-L$ space.
From \citet{ggs03}, by permission of Blackwell Publishing.
}
\end{figure*}


\section{Some Properties of Ultraluminous X-ray Sources}  \label{s:two}

The most X-ray luminous off-nucleus point-like objects in nearby galaxies
are refered to as Ultraluminous X-ray sources  or ULXs.
Their high X-ray luminosities ($L \ga 10^{39}$~\ergl\ in
the few tenths to $\sim$10~keV range) imply high accretor masses;
hence their interest as a possible evolutionary link
between stellar-mass black holes ($M_{\rm BH} \sim 1-20$~\msun)
known throughout the Local Group and super-massive black holes
 ($M_{\rm BH} \sim 10^6-10^9$~\msun) ubiquitous in the nuclei of galaxies.

Numerous 
 X-ray spectral and time-variability studies have shown that ULXs are,
 as a class, not steady thermal sources as expected from hot gas or SNRs 
 but must be some form of X-ray binary (XRB). 
The association of the most luminous sources with 
 galaxy-wide star-formation indicators, 
 such as FIR luminosity as discussed above,
 favor high-mass rather than low-mass XRBs. 

High-mass XRBs in our Galaxy and the Magellanic Clouds
 are those with short-lived, O or B type mass-donor stars.
If this is the case also for ULXs, then they should be found 
 preferentially in young star-forming regions such as \hii\ regions.
Population synthesis models have shown that the highest
X-ray luminosities, corresponding to the highest mass-transfer rates,
are expected for high-mass XRBs aged roughly 4 to 10~Myrs old when 
the mass-donor companion evolves off the main sequence.
However, unambigously identifying optical counterparts to ULXs
 requires monitoring campaigns at high spatial resolution using 
 {\it Hubble} or ground-based 8-m class telescopes.
More fruitful are studies of CMDs of objects in the neighborhood of ULXs,
 \citep[e.g.,][]{soria05}, which can narrow down the age of all 
likely counterparts in the region.

What is needed is
 a broadly-applicable method to easily quantify the star-forming
 properties of the local environments of ULXs without the expense of
 deep pointed obervations of individual ULXs.
We have recently completed a survey using such a method \citep{swartz09},
 calibrated on the SDSS optical colors of known \hii\ regions,
 to distinguish young, star-forming regions in nearby galaxies. 
We have used these criteria to examine the 
 100$\times$100~pc$^2$ regions around ULX candidates to determine
 if they are associated with the young stellar population.
ULX environments in our study are designated
 as ``star-forming'' if they are as blue as, or bluer
 than a typical \hii\ region in its host galaxy (taking also
 into account the reddening effects of dust and of old underlying
 stellar populations).
A useful feature of this method is that colors are distance-independent
 quantities, and are measured on distance-independent regions.

The 58 galaxies included in this study are a subset of a complete
sample of 140 galaxies selected for X-ray analysis of their ULX populations.
The X-ray sample galaxies are those 
 that are in both the set of 
 all galaxies within 15~Mpc contained in the UGC 
 with photographic magnitude $m_p<14.5$~mag and the set of all galaxies listed in the 
 {\sl IRAS} catalogs with a flux 
 $f_{\rm FIR}\ge10^{-13.3}$~\ergcms.
The 58 galaxies included in our study were those within the SDSS DR6 footprint
with inclination $i<65$\arcdeg.

We found that fully 60\% of those ULX regions 
 with sufficient signal ($21/35$) are in star-forming regions.
By definition, these regions are as blue or bluer than typical \hii\ regions 
 of their host galaxies and 
 therefore are likely of an age also typical of \hii\ regions which is $\la$10~Myr;
 the characteristic lifetime of the least-massive LyC-producing stars, i.e., late~O to
early~B stars of about 15 to 20~\msun. 

At the same time, we found that many of the most luminous ULXs in our sample
are located in faint or non-starforming regions. 
Some of these may be heavily reddened by dust.
However, we speculate that the most luminous ULXs
(or equivalently, the most common phases of very high mass transfer)
are biased towards early B-type donors with an initial mass of
$\sim$10$-$15~\msun\ and an age $\sim$10$-$20~Myr
(perhaps at the stage where the B star expands to become
a blue supergiant). In that case, we expect very little residual
H$\alpha$ emission from their surroundings, the O stars having
already evolved and died. 
Thus, we are beginning to constrain the age of the donor stars in
ULX binary systems. We hope to soon constrain the age and mass of 
the accretor.

\section{A Search for the Most Luminous ULXs} \label{s:three}

It was shown in \S\ref{s:one}
 that the X-ray luminosity function (XLF) of sources 
in the Milky Way and nearby normal spiral and starburst galaxies is a 
smooth power-law distribution from $\ll$10$^{39}$ up to 
$\sim$10$^{40}$~\ergl\ 
\citep{ggs03,swartz04,liu06}.
There are too few ULX candidates above $\sim$2$\times$10$^{40}$~\ergl\ to state definitively 
whether or not the smooth distribution continues to higher luminosities.
There are hints that a break or cutoff occurs at this point which, if real, imposes 
fundamental constraints on the physical properties of ULXs.

The most important of these constraints is the mass of the accretor. 
We know empirically, from studies of Galactic sources, that
the most reliable indirect mass indicator is the Eddington limit. 
If there is a cutoff in the luminosity distribution at $\sim$2$\times$10$^{40}$~\ergl,
then this suggests a maximum black hole mass of $\sim$100~\msun. 
The precise mass limit would depend on the degree of anisotropy in the emission
pattern \citep[e.g.,][]{king08} and on the degree of super-Eddington luminosity; 
perhaps allowing for a factor of 1 to 2 lower limiting mass. 
On the other hand, if ULX luminosities
extend beyond $\sim$5$\times$10$^{40}$~\ergl\ (or if the luminosity function 
flattens), then masses $>$100~\msun\ are 
indicated, at least for the brightest sources.

This is a critical dividing line. 
Theoretically, stellar cores up to $\sim$70~\msun\ can form equal mass black holes by
direct collapse \citep{yungelson08, heger03}
The final black hole mass may be even slightly larger if there is
residual fall-back of any remaining stellar envelope. Above this core mass, stars
disrupt via pair instability supernovae, leaving no remnant. 
Thus, something beyond normal single-star evolution will be required if 
ULXs are found to have luminosities much higher than a few 10$^{40}$~\ergl.


Figure~2 shows the {\sl differential} XLF for 
spiral galaxies taken from our original survey \citep{swartz04}. There are 
57 spiral galaxies in this sample and 97 ULX candidates but only a dozen
above $10^{40}$~\ergl. The fitted curves 
indicate a potential cutoff in the XLF at 2.07$\times$10$^{40}$~\ergl\ but the
change in the fit statistic corresponds to an improvement 
over a single power law model at only 85\% confidence.

\begin{figure*}[ht]
\begin{center}
\scalebox{0.4}{\rotatebox{-90}{\includegraphics{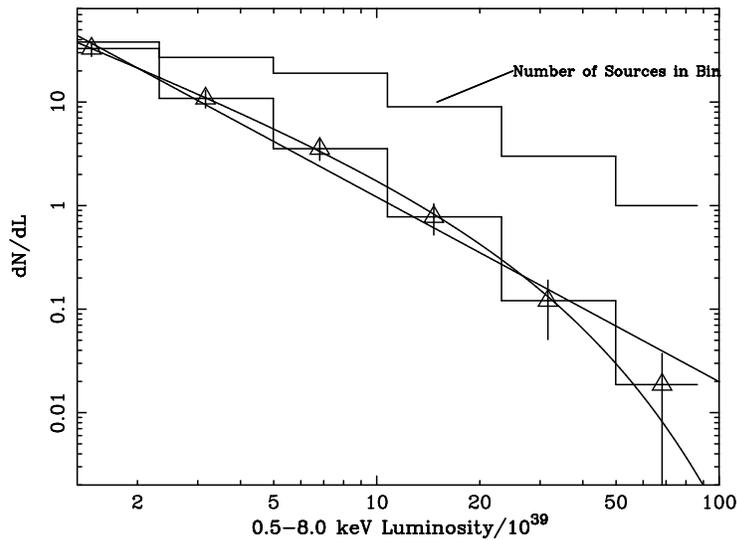}}}
\end{center}
\caption[]{Differential XLF for spiral galaxies taken from \citet{swartz04}.
Three log-linear bins were used per decade in intrinsic 0.5--8.0~keV 
luminosity for the 97 ULX candidates. 
The best fit power law model and best fit exponentially cutoff power law model
are shown.
The exponentially cutoff power law is a better fit at 85\% confidence.
The upper curve shows the actual number of objects in each bin.
}
\end{figure*}

We can estimate the number of ULXs needed to distinguish with high confidence
between the power law and exponentially cutoff power law from the trends 
in the existing data. Assuming that the 
exponential cutoff power law is the true functional form, we made 
a Monte Carlo simulation that computed the expected number of sources 
in each luminosity bin
then generated a random variable with mean equal to the expected number and with
a Poisson distribution. We then fit the simulated XLF with both models and compared
the difference in the (maximum-likelihood) fit statistic. We conclude from this that
about twice the current number of ULXs are needed to 
distinguish between these models at a $>$95\% confidence level.
This corresponds to $\sim$30 additional ULXs above a luminosity 
5$\times$10$^{39}$~\ergl.

Since both the {\sl number} and {\sl peak luminosity} 
of ULXs strongly correlates
with galaxy-wide SFR
and anti-correlates with nearest-neighbor distance, a logical 
choice of targets for our search are the high SFR
interacting galaxies. 
Happily,
\citet{smith07} have performed an in-depth study of the properties of a 
sample of pre-merger interacting galaxy pairs based on 
\sst\ luminosities and colors. They selected
isolated, tidally disturbed, binary systems
from the Atlas of Peculiar Galaxies \citep{arp66}
that are within $\sim$150 Mpc and that have 
large angular size to allow for good spatial resolution with \sst.
Of the 35 Arp pairs that met their criteria, we selected 
7 galaxy pairs for a snapshot X-ray survey with \cxo\ which we believe
will provide us with about 30 ULXs more luminous than 
5$\times$10$^{39}$~\ergl. 
These are being observed during the current \cxo\ observing round and the
results will be combined with archival data from several other galaxy
pairs in the \citet{smith07}.
So far, four Arp galaxy pairs have been observed and 
13 ULX candidates have been detected above 5$\times$10$^{39}$~\ergl.

\acknowledgements 
This research is supported in part by Chandra Award GO6-7081A
 issued by the Chandra X-ray Observatory Center which is operated by the 
 Smithsonian Astrophysical Observatory for and on behalf of NASA under 
 contract NAS8-03060.



\begin{thebibliography}{}
\bibitem[Arp(1966)]{arp66}Arp, H. 1966, Atlas of Peculiar Galaxies (Pasadena: Caltech)
\bibitem[Dalton \& Sarazin(1995)]{dalton95}Dalton, W. W., \& Sarazin, C. L. ApJ, 440, 280
\bibitem[David, Jones, \& Forman(1992)]{djf92}David, L. P., Jones, C., \& Forman, W. 1992, ApJ, 388, 82
\bibitem[Grimm \etal(2003)]{ggs03}Grimm, H. -J., Gilfanov, M. \& Sunyaev, R. 2003, MNRAS, 339, 793
\bibitem[Heger \etal(2003)]{heger03}Heger, A. , \etal\ 2003, ApJ, 591, 288
\bibitem[Helfand \& Moran(2001)]{helfand01}Helfand, D. J., \& Moran, E. C. ApJ, 554, 27
\bibitem[King(2008)]{king08}King, A. R. 2008, MNRAS, 385, L113
\bibitem[Liu \etal(2006)]{liu06}Liu, J. -F., Bregman, J. N., \& Irwin, J. 2006, ApJ, 642, 171
\bibitem[Smith \etal(2007)]{smith07}Smith, B. J., Struck, C., Hancock, M., Appleton, P. N., Charmandaris, V. \& Reach, W. T. 2007, AJ, 133, 791
\bibitem[Soria \etal(2005)]{soria05}Soria, R., Cropper, M., Pakull, M., Mushotzky, R. \& Wu, K. MNRAS, 356, 12
\bibitem[Swartz \etal(2004)]{swartz04}Swartz, D. A., Ghosh, K. K., Tennant, A. F., \& Wu, K. 2004, ApJS, 154, 519
\bibitem[Swartz \etal(2009)]{swartz09}Swartz, D. A., Tennant, A. F., \& Soria, R. 2009, ApJ, in press (arXiv:0907.4718)
\bibitem[Yungelson \etal(2008)]{yungelson08}Yungelson, L. R. \etal\ 2008, A\&A, 477, 223
\end{thebibliography}
\end{document}